# Current-voltage characteristics of semiconductor-coupled superconducting weak links with large electrode separations

Mason Thomas, Hans-Richard Blank, Ki C. Wong,
Herbert Kroemer,* and Evelyn Hu

*Department of Electrical and Computer Engineering,
and QUEST, Center for Quantized Electronic Structures,
University of California, Santa Barbara CA 93106*

We have studied the current-voltage characteristics of superconducting weak links in which the coupling medium is the 2-D electron gas in InAs-based semiconductor quantum wells, with relatively large (typically 0.5μm) separations between niobium electrodes. The devices exhibit Josephson-like current-voltage characteristics; however, the falloff of the differential resistance with decreasing temperature is thermally activated, and is orders of magnitude slower than for more conventional weak links. Most unexpectedly, the thermal activation energies are found to be proportional to the width of the device, taken perpendicular to the current flow. This behavior falls outside the range of established theories; we propose that it is a fluctuation effect caused by giant shot noise associated with multiple Andreev reflections. The possibility of non-equilibrium effects is discussed.

## I. INTRODUCTION

There has recently been some interest in the electrical properties of superconducting weak links that use, as the coupling medium between the two superconducting electrodes, the high-mobility quasi-two-dimensional electron gas in a GaAs- or InAs-based semiconductor heterostructure.[1-16] Much of that interest was stimulated by the 1978/1980 papers of Silver et al.[17] and Clark et al.,[18] who proposed three-terminal gate-modulatable weak-link devices, referred to as *hybrid Josephson field effect transistors*, or briefly JOFETs. These devices draw on the ability to modulate the electron concentration in a thin semiconductor layer via a gate electrode, and thereby modulate the Josephson critical current. The authors recognized the need for very thin semiconductor coupling layers with long electron mean free paths, a combination naturally leading to the investigation of the high-mobility quasi-two-dimensional electron gas in suitable semiconductor heterostructures. Silver et al. also recognized the specific advantages of InAs as the semiconductor: The Fermi level at metal-to-InAs interfaces falls inside the InAs conduction band,[19-21] thus leading to a freedom from Schottky barriers that would impede the flow of electrons. Most of the subsequent work has indeed used InAs or (In,Ga)As in various configurations; important exceptions are the work of Ivanov et al.,[1,2] and more recently of Marsh et al.,[7] who employed GaAs-based heterostructures with (superconducting) indium alloy contacts, another combination with very low interface barriers approaching the barrier properties on InAs.

In addition to the use of true heterostructures, some investigators have employed, as the coupling medium, the 2-D electron gas (2DEG) in the n-type surface inversion layer naturally present on bulk p-type InAs,[17, 22-24] a possibility already recognized by Silver et al. The only truly successful work of this kind was that by Chrestin and Merkt,[24] who recently used Nb electrodes with separations down to 20 nm, at which point the inherent limitations of the inversion layer scheme (low electron sheet concentrations and strong surface scattering) are largely overcome by the proximity of the electrodes.

In the present paper, we study the current-voltage characteristics (CVCs) of two-terminal devices in which the electrode separation $L$ is larger than the coherence length $\xi_n$ in the 2-D semiconductor, but still much smaller than the elastic mean free path $\lambda_{el}$ of the electrons (ballistic rather than diffusive limit). In this regime, strong superconducting coupling effects persist, largely as a result of the strong multiple phase-coherent Andreev reflections (ARs) that take place at the super-semi interfaces. Our rationale for interest in this regime goes beyond the scientific question of what happens when the electrodes are pulled farther apart. For device applications, working with larger separations offers

       

potential advantages: (a) A reduction of the parasitic inter-electrode capacitance that is electrically in parallel with the intrinsic Josephson junction. This would not only improve the high-frequency properties of the weak links, but also reduce undesirable hysteresis effects in the CVCs. (b) Larger electrode separations should make it easier to achieve tight tolerances on the *relative* fluctuations $\Delta L/L$ in the separation, and hence on device-to-device fluctuations in the critical currents, an important limitation in large-scale integrated Josephson circuits.

The existing literature on 2DEG-coupled weak links with large electrode separations is sparse: The most impressive experimental data in that category published so far are those by Marsh et al.,[7] who report Josephson-like CVCs at 1.6K for a GaAs-based weak link with an inter-electrode separation of 1μm, but a (clean-limit) coherence length at 1.6K of only about 0.24μm. These observations are strongly supported by some of our own work, which falls into a similar parameter range, although we did not stress the relative magnitudes of electrode separation and coherence length.[4-6, 11-15]

On the theoretical side, the work most relevant to the devices of interest here is probably that by Kümmel, Gunsenheimer, Nicolsky, Zaikin, et al. (KGNZ),[25-27] (where additional references to earlier work can be found). These authors studied the operation of weak links in the ballistic limit, emphasizing the central role of multiple Andreev reflections in such devices. Although their work is not explicitly directed towards devices with electrode separations larger than the coherence length, that case is implicitly contained in their work. More recently, the case $L > \xi_n$ has drawn some attention by others,[28, 29] but in a form that does not lend itself readily to a comparison with experimental data on ballistic devices. In fact, none of the theoretical papers cited make significant contact with experiment. This is presumably the result of the shortage of relevant experimental work. One of the purposes of the present paper is to help filling this gap. A second purpose is to point out gaps in the theoretical understanding of the observations.

Our principal interest is in the CVCs of devices in the regime $\xi_n < L < \lambda_{el}$ especially in their residual resistance at low bias, and in the temperature dependences of the low-bias regime. To enhance the measurement sensitivity in the limit of near-vanishing resistance, all measurements reported here were made on series arrays of a large number of individual weak links (typically 300), in a grating geometry introduced by Nguyen,[30] shown schematically in Fig. 1.

At temperatures above the critical temperature $T_{Nb}$ of the Nb electrodes (8.5 ± 0.5 K, depending on purity), the CVCs of all samples are perfectly linear; they resemble a barrier-free normal resistor with a resistance that is only weakly temperature-dependent, reflecting the weakly temperature-dependent mobilities and the essentially temperature-independent electron concentrations in the highly degenerate semiconductor.

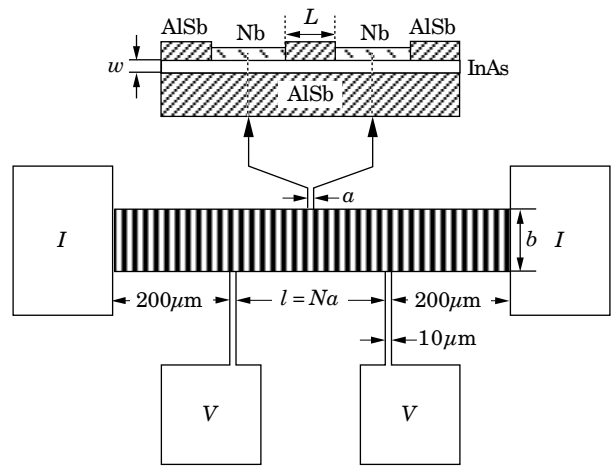

FIG. 1. Series arrays used for studying the low-bias resistance of the weak links, made by molecular beam epitaxy and laser holography.[6, 11] The top shows a cross-section through a pair of Nb lines separated by a narrow stripe of InAs-AlSb quantum well, the bottom shows the overall layout. All I-V measurements are four-point measurements, made by imposing a current $I$ via the outer contacts and measuring the voltage $V$ between the inner contacts. In all samples reported here, the InAs quantum well width $w$ was 15 nm.

As soon as the temperature falls below $T_{Nb}$, a non-linearity develops at zero bias, with a *decreased* differential resistance. Almost all of this non-linearity occurs over a voltage range that is narrow compared to $kT/e$, suggesting a collective phenomenon rather than single-electron behavior. With decreasing temperature, the zero-bias resistance $R_0 = dV/dI|_{I=0}$ decreases, usually over several orders of magnitude. In "good" samples, $R_0$ eventually drops below the noise floor of our measurement setup, around 0.003 Ω (about 10 μΩ per cell). There is never a discontinuity in $R_0$, and even in the arrays with the steepest falloff of $R_0$, the range of falloff is several Kelvin wide. This behavior is drastically different from that of Josephson tunnel junctions and more conventional (short) weak links, where the falloff of the zero-bias resistance is much steeper, typically over temperature ranges on the order of a few tens of millikelvin.

Fig. 2 gives an Arrhenius plot of $R_0$ for one of our best samples (sample A; see Table I below), measured directly with a small (2μA) dithering current. The plot indicates a thermally activated behavior with an *apparent* activation energy of about 24meV. A *qualitatively* similar behavior is shown by most samples, but with large *quantitative* sample-to-sample variations, to be discussed later. We call the activation energies thus obtained *apparent* activation energies, because the slope in an Arrhenius plot is not equal to the true activation energy when the latter depends itself on temperature. In the specific case of an Arrhenius plot showing a well-defined straight line, as in Fig. 2, the slope represents the activation energy linearly *extrapolated* to $T = 0$, which in turn need not coincide with the true activation energy *at* $T = 0$. With the true activation energy almost certainly decreasing with increasing temperature, its value in the temperature range shown in Fig. 2 must be lower than 24 meV, possibly much lower. Its exact determination requires information not contained in the Arrhenius plot itself; we will return to that point later.

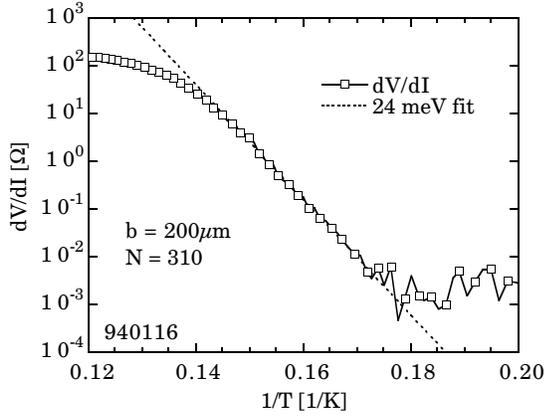

FIG. 2. Arrhenius plot of the zero-bias resistance $R_0 = \Delta V/\Delta I$ of sample A, measured with a small a.c. dithering current $\Delta I$ of 2 μA at 497 Hz (from ref. 14). The nearly-straight-line portion corresponds to an *apparent* activation energy of 24 meV. The leveling-out around 0.003 Ω represents the noise floor of our experimental setup, rather than a true limiting resistance. The array has 310 individual devices; a width (perpendicular to the current flow) of 200 μm; a period of 0.96 μm, and an inter-electrode separation of about 0.5 μm.

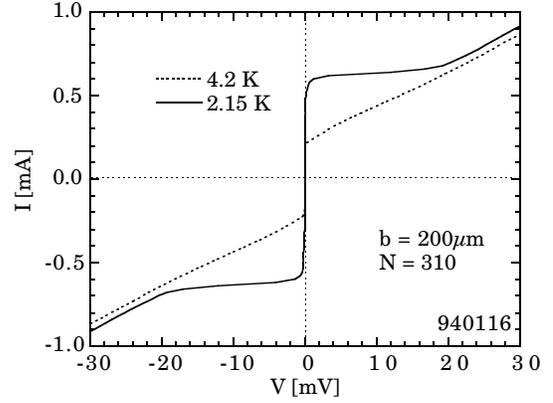

FIG. 3. Current-voltage characteristics at 2.15 K and 4.2 K of the same sample as in Fig. 2 (sample A).

Throughout the entire temperature range from $T_{Nb}$ down to the temperature where $R_0$ disappears below the noise floor, we are evidently dealing with an intermediate regime in which some aspects of superconductivity manifest themselves, but the devices are not yet in a fully-developed superconducting state. The exploration of this intermediate regime was one of the original motivations for the present work. A surprising discovery that was made in the course of the work was that the apparent activation energies of the zero-bias resistance, and by implication, the true activation energies, show a proportionality to the widths of the devices, that is, to the dimension perpendicular to the current flow (dimension $b$ in Fig. 1). This is the central topic of the present paper; it will be discussed in detail, in sec. III, along with other peculiar aspects of the width dependence.

Hand-in-hand with the decrease of the zero-bias resistance with decreasing temperature goes an upward shift of the entire CVC for bias voltages exceeding $kT/e$, a shift commonly referred to as *excess current*. Fig. 3 displays the resulting characteristics for sample A at 4.2 K and 2.15 K. The 4.2 K data almost appear as if the CVC consisted of a set of three straight lines, connected via two sharp corners that are displaced from the abscissa by the excess current. At 2.15 K, the (somewhat rounded) corners have shifted to higher currents, and are followed by voltage regions inside which the slope has noticeably flattened, eventually merging into almost the same asymptotes as at 4.2 K. (Except for the greatly expanded voltage scale, our 2.15 K characteristic is qualitatively very similar to the 1.6 K characteristics reported by Marsh et al.[7] for their single-gap GaAs device mentioned earlier.) Although in Fig. 3 these developments are shown at low temperatures at which the zero-bias resistance has dropped far below the noise floor, many devices show similar characteristics even while $R_0$ is still well above the noise floor. (See, for example, Fig. 6 below.)

The low-temperature characteristics superficially resemble those of (overdamped) Josephson junctions and more conventional (short) weak links, with a current at the corners of the CVC presumably being the Josephson critical current $I_J$. However, there are pronounced differences, the excess current being the most obvious. Most samples also exhibit, superimposed on the quasi-asymptotic resistance, the weak oscillatory subharmonic gap structure that serves as the "fingerprint" of multiple Andreev reflections.[31] It is not readily visible in "straight" CVCs such as Fig. 3, but shows up clearly as an oscillation in the differential characteristics. We shall see later that the identification of the Josephson critical current with the current at the CVC corners is open to question; hence we will refer to the latter here as *corner current*.

Qualitatively, the overall behavior described above has been observed regularly in all samples in which the resistance dropped sufficiently rapidly below the noise floor, and it may therefore be considered "typical" in a qualitative sense. However, there are major *quantitative* sample-to-sample variations, both in the magnitudes of the currents observed, and in the exact shapes and temperature dependences of the CVCs. Despite these variations, the occurrence of an excess current and the subharmonic gap structure at sufficiently low temperatures has been a common phenomenon in all samples.

The rest of the paper is organized as follows. Section II gives relevant materials and sample parameters. Section III discusses the width dependence of the CVCs in the *intermediate* temperature range, defined as the range below the critical temperature $T_{Nb}$ of the Nb electrodes, but at temperatures at which the zero-bias differential resistance has not yet dropped below the noise floor. Section IV discusses certain non-equilibrium aspects of the current flow near and above the CVC corners.

## II. MATERIALS AND SAMPLE PARAMETERS

All samples reported here were based on 15nm-wide InAs quantum wells with AlSb or (Al,Ga)Sb barriers, grown by molecular beam epitaxy (MBE) and modulation-doped with Te. All electrode separations were approximately 0.5μm. Technological details, including the modulation doping technique employed, can be found in refs. 32-34. Table I lists the electron sheet concentrations and mobilities for the selected samples reported here. The values given represent measurements on the original quantum well wafers before further processing. As we will discuss below, they do not necessarily reflect the values present underneath the Nb stripes after processing.

The mobilities in the material from which our samples are prepared are such that the samples are safely in the clean limit. We express the clean-limit coherence lengths in the normal region in the form $\xi_{nc} = \hbar v_F / 2\pi k_B T$, and the elastic mean free paths in the form $\lambda_{el} = \mu \cdot \hbar k_F / e$. Here, $v_F$ is the Fermi velocity, $\mu$ the electron mobility, and $k_F = \sqrt{2\pi N_s}$ the Fermi wave number. These forms remain valid in the presence of the strong non-parabolicity that exists in the conduction band of degenerately doped InAs. In InAs, the Fermi velocity saturates with increasing doping at about $1.2 \times 10^8$ cm/s. At the doping levels employed, a significant fraction of the electrons always spills over into the second 2-D subband,[33] where the Fermi velocity is somewhat lower. Assuming a value of $1 \times 10^8$ cm/s as a realistic upper limit, we estimate a coherence length at 4.2 K of at most 0.29 μm, safely below the electrode separation $L$, and exceeding the latter only for temperatures below about 2.4 K. On the other hand, the elastic mean free paths $\lambda_{el}$ tend to be much larger than the electrode separations $L$: For a mobility $\mu$ of, say, $10^5$ cm$^2$V$^{-1}$s$^{-1}$, and an electron sheet concentration $N_s$ of $5 \times 10^{12}$ cm$^{-2}$, we have $\lambda_{el} \approx 3.7$ μm, well above all our electrode separations, placing the samples safely in the ballistic rather than diffusive regime. What ultimately matters more than the elastic mean free path is the inelastic mean free path, which is much larger, but the exact value of which is unknown.

As pointed out in the *Introduction*, much of our work was plagued by strong sample-to-sample variations, which manifested themselves in several ways. The largest variations were in the apparent activation energies of the zero-bias resistance, which, at the present stage of the technology, still fluctuate over a range on the order of 2:1 for supposedly identical samples (The 24 meV of Fig. 2 is one of the largest values we have observed). Also, the well-defined linear range in Fig. 2 should not be taken for granted. In many samples the slope increased continuously with decreasing temperature. In a few samples the linear range was preceded at its high-temperature end by a narrow region of *steeper* slope just below the $T_{Nb}$ of the electrodes (see, for example, Fig. 2 in ref. 13).

Ignoring effects of sample geometry, which will be discussed separately, the sample-to-sample variations can be traced to several sources:

(a) The dominant (and least well-controlled) source are processing-induced variations in the "quality" of the Nb-InAs interfaces. The exact nature of these variations is not clear; Magnee et al., in similar studies of InAs-Nb interfaces,[35] report damage during sputter cleaning of the interface, but we find such variations also in samples in which sputter cleaning was not used. One of the manifestations of processing-dependence is an increase in the sheet resistance of the samples after processing, by a sample-dependent amount, relative to the values measured on the InAs quantum wells beforehand. Recent work by den Hartog et al.[36] indicates that the processing can change the scattering at the Nb/InAs interface from specular to diffusive, which could explain the observations. Inasmuch as we are principally interested in the transport through the semiconductor *between* the Nb stripes, where the properties

Table I. Relevant parameters for the selected samples for which measurements are presented in this paper. The samples are listed by MBE growth (col. 2). Samples given a common letter in col. 1 (Such as C1 through C4) were co-processed. Col. 3 indicates the electron sheet concentration, col. 4 the low-temperature (≈10K) electron mobility, col. 5 the array width $b$, and col. 6 the separation $l$ between the voltage electrodes.

| 1 | 2 | 3 | 4 | 5 | 6 |
|---|---|---|---|---|---|
|   | MBE run ID | $N_s$ cm$^{-2}$ | $\mu$ cm$^2$/Vs | $b$ μm | $l$ μm |
| A | 940116 | $8.5 \times 10^{12}$ | 89,000 | 200 | 300 |
| B |        |                      |        | 95  | 300 |
| C1 | 950531 | $5.5 \times 10^{12}$ | 222,000 | 90 | 300 |
| C2 |        |                      |        | 90  | 425 |
| C3 |        |                      |        | 65  | 300 |
| C4 |        |                      |        | 65  | 425 |
| D1 | 970813 | $8.7 \times 10^{12}$ | 67,000 | 150 | 200 |
| D2 |        |                      |        | 100 | 200 |

should not have changed, we have not followed up these processing effects in the present work.

(b) Data on samples with (deliberately) different MBE growth parameters indicate that the apparent activation energies increase with increasing electron sheet concentration $N$ in the 2-D electron gas, at least over the limited range studied, from about $5.5 \times 10^{12}\,\text{cm}^{-2}$ to about $8.5 \times 10^{12}\,\text{cm}^{-2}$. However, we do not have enough data to quantify this dependence and separate it from the effects of the uncontrolled interface variations mentioned under (a).

(c) Finally, sample-to-sample variations may be mimicked by uncontrolled fluctuations in the measurement environment, due to stray magnetic fields, or to noise injection via the measurement leads. To suppress the strong effects of even weak magnetic fields, documented in earlier work of ours,[11] all measurements reported here were made inside a superconducting Pb shield surrounding the sample. To suppress noise injection via the measurement leads, low-pass filters were inserted into all electrical lines leading into the cryostat, containing damping resistors that were themselves located inside the low-temperature measurement space.

## III. WIDTH DEPENDENCE

### A. Experimental Results

A naive argument would suggest that the zero-bias resistance should scale inversely proportionally to the sample width $b$, which would imply a vertical shift on an Arrhenius plot, but not a change in the slope of the plot, and hence not a change in the (apparent) activation energy. Although this is what we observe at temperatures above $T_{\text{Nb}}$, we find an unexpected dependence of the Arrhenius slope on width as soon as the temperature is dropped below $T_{\text{Nb}}$, with apparent activation energies proportional to $b$. Put differently: A 200µm-wide device behaves altogether differently from two 100µm-wide devices connected in parallel; with decreasing temperature, the resistance of a 200µm-wide device falls off roughly twice as rapidly (on a log scale!) as that of a 100µm-wide one. As we stated in the *Introduction*, this is the central topic of the present paper.

Such a dependence was first noticed as a byproduct of a study of magnetic flux quantization effects in our arrays.[11] During that study, a sample with an initial width of 98µm was etched down, first to a width of 44µm, then to 16µm. The devices showed the expected dependence of flux quantization effects on width, but also an unexpected decrease in the apparent activation energies of their zero-bias resistance.[37] (Those data were not included in ref. 11)

The virtue of this etch-down method is that it does not change the remaining super-semi interface, eliminating the dominant source of uncontrolled sample-to-sample variations. However, the resulting geometry is hard to control, and in the work reported here we have taken a different approach. Arrays with different outer dimensions were manufactured on a common wafer, using a common holographic exposure, and subsequently co-processed to ensure identically-processed Nb-to-InAs interfaces. Only those final processing steps were performed separately that define the array widths and the placements of the voltage electrodes.

Fig. 4 shows the results for a set of four samples (C1 through C4) that were prepared in this way. Two widths (65 µm and 90 µm) and two lengths between the voltage probes (300µm and 425µm) were employed, giving four combinations of length and width. It is evident from Fig. 4 that, in the thermally-activated temperature range (below ~8K), the apparent activation energies are independent of the length (and hence of the number of Nb-InAs junctions in series), but depend on the width. Within the accuracy with which the activation energies can be extracted from the imperfect Arrhenius plots, the extracted values, 3.6meV and 2.6meV, vary in the same ratio as the widths of the arrays, $E/b \approx 0.04\,\text{meV}/\mu\text{m}$, extrapolating to zero activation energy for zero width.

Note that this $E/b$ ratio is much smaller than for sample A, where $E/b \approx 0.12\,\text{meV}/\mu\text{m}$. Much of the difference is believed to be due to the higher electron concentration in the MBE growth from which sample A was prepared; some of it may reflect post-growth processing differences. Relative to sample A, the interface quality of the four samples of Fig. 4 is of course again subject to process variations, but at least the four interfaces should share a common quality, whatever it may be.

To check the width-proportionality of the apparent activation energy for higher doping levels, a second pair of co-processed samples was prepared, at the opposite end of the doping range (D1 and D2 in Table I). They, too, exhibited this behavior, with $E$-values of 24 meV and 16 meV, corresponding to a ratio $E/b \approx 0.16$ meV/μm, higher even than for sample A. Data on other samples, although taken under less stringently controlled conditions, but over a wider range of parameters, are all in good agreement with a width-proportional activation energy, at least over the range of widths we have explored, from 10 μm to 200 μm.[37]

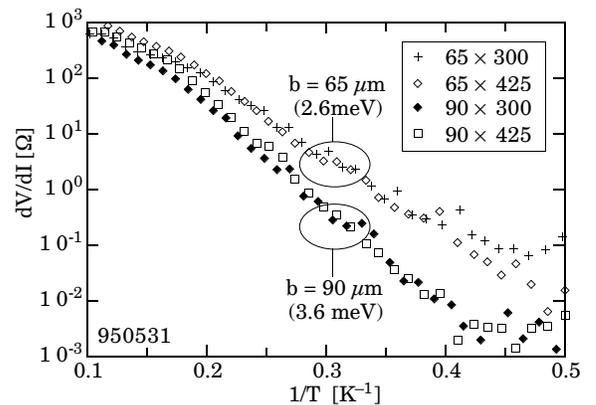

FIG. 4. Arrhenius plot of the zero-bias differential resistance of four samples with widths of 65 and 90µm, and lengths of 300 and 425µm (C1 through C4 in Table I). The two samples with a width of 65µm have the same activation energy—2.6meV—independent of length. Similarly, the two samples with a width of 90µm have a common activation energy of 3.6meV. From ref. 37.

Measurements of the zero-bias resistance give only limited information about the temperature dependence of the CVCs, and to complement the data of Fig. 4. we have also measured the full $R(I)$ characteristics at 2 K and 4 K for the samples shown there. Fig. 5 displays the 2 K results for those two arrays that had a common length of 425 μm, but different widths (the data for the 300 μm-long pair are very similar). If the resistances would simply scale inversely proportional to the array widths, this would show on the semi-log plot of Fig. 5 as a small *vertical* displacement of the two curves relative to each other. Although such a small displacement can be seen in the high-current range, where the resistances have essentially saturated with increasing current and reached an asymptotic value $R_a$, there is a much more pronounced *horizontal* shift: Above the noise floor, the curve for the 90 μm-wide sample is shifted *as a whole* to higher (absolute) currents by about 30 μA.

It is useful to re-formulate this peculiar behavior as follows. We define a *characteristic current $I_c$* for each curve, associated with a readily identifiable feature just above the CVC corner, namely, the end of the sharp overshoot peak of $R(I)$, the point where $R(I)$ has returned to the asymptotic resistance $R_a$. This characteristic current is, to a high accuracy, proportional to the device width, as expected. The behavior of the normalized resistance $R/R_a$ for currents below $I_c$ may then be expressed as a law of the general form

$$R(I)/R_a = f(I - I_c), \qquad (1)$$

where the function $f$ does not depend *explicitly* on the sample width, only $I_c$ does. Put differently, for samples that have different widths (yet are otherwise identical), the resistance depends only on the amount $\delta I = I_c - I$ by which the actual current $I$ is still below the characteristic current $I_c$; they exhibit the same resistance at the same current difference $\delta I$. We believe that this simple functional dependence provides an essential clue as to the mechanism of this resistance, to be discussed below.

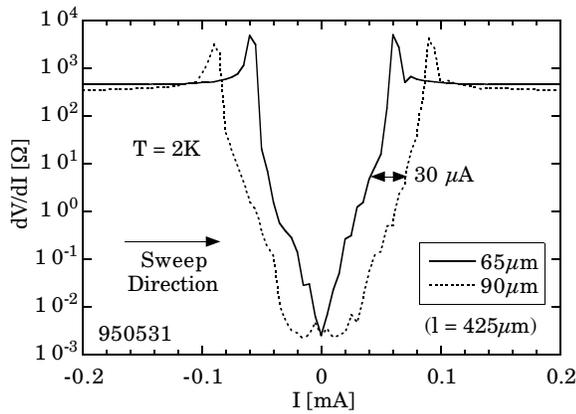

FIG. 5. Differential resistance vs. current at 2 K for two co-processed arrays of different widths (90 and 65 μm), but a common length (425 μm). Above the noise floor, the two curves are essentially identical, except for a large horizontal displacement by about 30 μA (and a very small vertical shift reflecting the difference in array width). The data for the 300 μm-long arrays look very similar.

The function $f$ is of course still a function of such other sample parameters as the interface quality, the electron sheet concentration, and especially of the temperature. For resistance values more than a decade below the asymptotic value $R_a$, the data can be fitted, to a good approximation, to a simple exponential, which may be written

$$R(I)/R_a = \exp\left[(I - I_c)/I_s\right], \qquad (2)$$

where $I_s$ is a width-independent measure of the steepness of the exponential rise, approximately 5.4 μA for this particular pair of samples at 2 K.

Setting $I = 0$ in (2) yields a zero-bias resistance that decreases exponentially with the corner current,

$$R(0)/R_a = \exp\left[-I_c/I_s\right], \qquad (3)$$

The scaling current $I_s$ is of course temperature-dependent; the Arrhenius plots of Fig. 4 suggest that $I_s$ is roughly proportional to the temperature $T$. We may express this proportionality in the form

$$I_s = \gamma \cdot e k_B T / \hbar. \qquad (4)$$

where $\gamma$ is an alternative dimensionless fitting parameter, which is typically on the order 100 or more. Inserting (4) into (3) leads to the final form

$$R(0)/R_a = \exp\left[-\hbar I_c / \gamma e k_B T\right]. \qquad (5)$$

Because $I_c$ is proportional to the device width, this result implies a width dependence of the activation energy for the zero-bias resistance.

### B. Interpretation

The behavior described above suggests that the finite differential resistance below the characteristic current is the result of statistical current fluctuations, *qualitatively* similar to the way thermal current fluctuations cause a nonzero residual resistance to occur already in Josephson tunnel junctions just below their critical current, as described by the theory of Ambegaokar and Halperin (AH)[38] (see also Sec. 6.3.3 of Tinkham[39]).

More specifically, the functional relation (1) suggests that $I_c$ acts as a *threshold current* for the appearance of a voltage, similar to the way the Josephson critical current $I_J$ in tunnel junctions acts as a threshold current. In the absence of statistical noise fluctuations the devices would presumably have zero differential resistance below that threshold. However, fluctuations of the applied bias current cause voltage pulses to appear at the device terminals whenever the current exceeds the threshold. The time average over voltage pulses yields a d.c. voltage; its magnitude increases with increasing d.c. current; the derivative $dV/dI$ of this averaged voltage is the observed differential resistance.

In terms of such a fluctuation model, the current $I_s$ introduced in (3) would have to be interpreted as a measure of the amplitude distribution of the fluctuations; the exponential law (2) suggests that the probability per unit time of a current fluctuation exceeding $\delta I$ falls off exponentially with increasing $\delta I$, with a $1/e$-parameter $I_s$.

$$p(\delta I) = p_0 \cdot \exp(-\delta I / I_s), \quad (6)$$

where $p_0$ is an extrapolated value for the limit $\delta I = 0$. Our data require an $I_s$-value independent of the width $b$. This does *not* imply that the fluctuation probability itself is width-independent; it simply means that the $b$-dependence—which must be expected to be present—is contained within the pre-factor $p_0$. In fact, we would expect that $p_0$ increases with width, presumably proportional to $\sqrt{\delta I}$. In terms of Fig. 5, the latter dependence would show up as a small vertical displacement of the $R(I)$ curves for different widths, similar to the small vertical shift at high currents, but presumably even smaller, because the fluctuations should be expected to add sub-linearly with increasing width. The precision of our data is insufficient to separate this weak dependence from the much stronger exponential dependence.

In the AH model, the current fluctuations are assumed to be simply the thermal-equilibrium Johnson noise that is associated with the shunting resistance in the familiar RSJ model[39] of these devices. In the limit $\hbar I_J \gg ek_BT$, the AH theory leads to a simple relation between zero-bias resistance and critical current of the form

$$R(0)/R_n \propto (\hbar I_J / ek_BT) \cdot \exp[-\hbar I_J / ek_BT], \quad (7)$$

where $R_n$ is the normal resistance of the junction, which serves as the shunting resistance in the RSJ model. For large values of the ratio $\hbar I_J / ek_BT$, the exponential dominates the temperature dependence. With $I_J$ presumably being proportional to the width of the current-carrying path, the AH theory, or any similar fluctuation-based theory, would provide a natural explanation for the width-dependence of our activation energies. Except for the factor preceding the exponential in (6) and the coefficient $\gamma$ in (5), the two expressions have a remarkable similarity, and it is in fact this similarity that suggests that our observed behavior is a fluctuation phenomenon, too.

However, in its existing form, the AH theory itself cannot be invoked as a quantitatively valid theory for our case. To demonstrate the magnitude of its failure, we draw on the fact that the AH theory gives not just a relation between $I_J$ and $R_0$, but predicts the entire CVC. Consider the samples of Fig. 5. If we plot the CVC for the 65 μm wide sample, we obtain the 2 K curve of Fig. 6. We have singled out the 65 μm sample because its zero-bias resistance even at 2 K is still above the noise floor, thereby permitting a test of the theory over the full current range. The shape of the CVC differs significantly from that predicted by the AH theory, but if we ignore this discrepancy and simply "force-fit" the data to the theory, this requires a critical current that is within at most ±10% of 55 μA. According to AH, this would correspond to an activation energy of about 226 meV, a factor 87 above the *apparent* activation energy extracted in Fig. 4 from the Arrhenius plot of $R_0$ for this device. Inasmuch as the apparent activation energy is the activation energy linearly extrapolated to $T = 0$, the true activation energy at $T = 2$ K must be even lower, potentially much lower, and the discrepancy must be even larger. Similar discrepancies have been found in all samples for which we have enough data to permit the comparison. In fact, the phenomenological fitting parameter $\gamma$ in (5) is a direct measure of the magnitude of the discrepancy.

This inapplicability of the AH theory is not surprising. The underlying RSJ model of Josephson tunnel junctions[39]

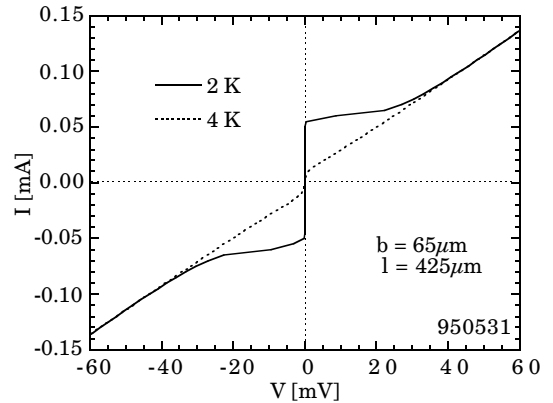

FIG. 6. Current-voltage characteristics of the 65 μm-wide sample of Fig. 5, at 2 K and 4 K. Even at 2 K, $R(0)$ for this sample has not yet dropped below the noise floor, yet the CVC shows very pronounced and remarkably sharp corners.

is based on the fact that the current in those devices can be cleanly separated into two physically distinct parts: An ideal (resistance- and noiseless) Josephson supercurrent due to pair tunneling, and a normal (resistive and noisy) current due to quasiparticle tunneling. It is the Johnson noise of the latter that causes the total current to fluctuate above the critical current of the former, and hence causes a nonzero residual resistance to occur. However, such a separation cannot be made for ballistic weak links in the intermediate temperature range covered here, where the behavior represents a far more complex hybrid between a conventional Josephson junction and a normal resistor.

Experimentally, the zero-bias resistance, although reduced, is still finite, arguing against a non-zero true Josephson critical current. Yet, we are clearly no longer dealing with a normal conductor. In particular, as was shown by Drexler et al.,[13] the a.c. Josephson effect, as manifested by pronounced Shapiro steps, persists up to temperatures at which the d.c. Josephson effect has all but disappeared.

Theoretically, the phase coupling between the two widely separated superconducting banks is not by Cooper pair tunneling, but by phase-coherent multiple Andreev reflections. It is known that, in the absence of competing processes, an unbroken sequence of ARs would indeed lead to a resistanceless supercurrent with a well-defined critical current.[40] But in reality, the AR sequence will, at random intervals, be interrupted by other scattering processes, including specifically *normal* reflections at the super-semi interfaces. As has been pointed out by Averin and Imam,[41] such interruptions lead to what the authors call "giant shot noise" in the supercurrent itself, and we believe that such a model is a more promising basis for the theory of the intermediate temperature range of ballistic weak links than the AH-modified RSJ model.

The reason why multiple ARs in ballistic weak links contain, within themselves, a mechanism for large current fluctuations, is easily understood by the following elementary argument. Consider the highly oversimplified model of an ideal ballistic weak link in which all scattering mechanisms other than scattering at the super-semi interfaces are ignored. Assume further that the probability for an AR at the interfaces is much higher than for a normal reflection. This means that the electrons will undergo trains of successive multiple ARs at alternating interfaces, interrupted, at irregular intervals, by a normal reflection. A train of *n* pairs of successive ARs (at opposite electrodes) causes *n* Cooper pairs to be moved from one of the superconducting banks to the other. A single intervening normal reflection will not itself cause any current to flow across the interface, but it will reverse the direction of current flow during the next train of multiple ARs! This is evidently a mechanism for large current fluctuations.

In their paper, Avery and Imam discuss the effects of these giant shot noise fluctuations on the noise properties of quantum *point* contacts.[41] Although in wide devices such as ours, such fluctuations would be much smaller *relative* to the d.c. current than for point contacts, we believe that a theory of a fluctuation-induced d.c. resistance based on this mechanism is a promising approach towards understanding the resistance properties of weak links such as ours. Working out the details of such a proposed model would go far beyond the primarily experimental scope of the present paper (and our expertise), and we have not attempted such a development.

## IV. NON-EQUILIBRIUM EFFECTS

In addition to noise fluctuations, a transport mechanism based on multiple ARs also suggests that the corner currents of the CVCs are not necessarily equilibrium Josephson critical currents, but may involve a substantial non-equilibrium effect. What is meant by this is the following.

As mentioned in the *Introduction*, the theory of weak links operating in the ballistic limit is fairly well developed (see KGNZ,[25-27] where additional references to earlier work can be found). Although KGNZ do not include fluctuation effects, their theory still represents a valid description of the basic dynamics of ballistic electrons in a normal conductor bounded by two superconductors, at least in the limit that the electron mean free path is sufficiently long. Scattering within the semiconductor is then weak (ballistic or "clean" limit), and the dominant scattering process is multiple phase-coherent Andreev scattering at the two superconductor-semiconductor interfaces. These assumptions should be satisfied, to a reasonable approximation, in the structures investigated by us. In fact, in another paper by the same group, the theory is explicitly applied to the Nb-InAs-Nb system.[42]

It is shown in the KGNZ theory that, under these conditions, and in the presence of a small but nonzero voltage bias, phase-coherent multiple ARs can lead to an energy distribution of the electrons in the semiconductor that differs strongly from a thermal-equilibrium Fermi distribution. In the limit of small bias voltages, average electron excess energies may become large compared to the bias energy $e \cdot V$, and may reach values on the order of a significant fraction of the superconducting gap parameter $\Delta$. But his implies that the conditions near the corners of the CVC are non-equilibrium conditions, and that the corner currents should not be expected to be purely equilibrium Josephson critical currents. In fact, in the limit of sufficiently long inelastic mean free paths, KGNZ predict CVCs that are remarkably similar to the 4.2 K characteristics shown in Fig. 3, with a very steep rise at very low voltages (in the microvolt range), terminated in a rounded corner that connects the steep low-voltage characteristics to an asymptotic characteristic with a slope that is essentially the same as above the Nb critical temperature, but offset by a finite excess current (see, for example, Fig. 1 in ref. 27). Furthermore, the corner current can be large compared to the true Josephson critical current, and much of the KGNZ theory remains applicable when the *true* Josephson critical current becomes arbitrarily small. We do not wish to claim that these conditions are met for the case of the 4.2 K data of Fig. 3, but one certainly should not take it for granted that the corner currents in Fig. 3 are ordinary equilibrium Josephson critical currents.

This idea that the currents near the CVC corners might contain a large non-equilibrium component, is strongly supported by observations of an anomalous a.c. Josephson effect in arrays of the kind discussed here. It was found by Drexler et al.[13] that such structures exhibit Shapiro steps not only at the expected voltage $V_J = Nh\nu/2e$, but similarly strong steps at one-half that voltage. Most significantly, with increasing temperature, the steps at $V_J$ decrease more rapidly than the half-voltage steps, and at sufficiently high temperatures, the latter dominate, persisting up to temperatures (7.5-8K) at which the d.c. characteristics shows no remaining evidence of superconductivity other than the slight enhancement of the zero-bias resistance mentioned in the *Introduction*. More recent observations by Lehnert et al.[43] on a single-gap device have corroborated these observations, and have provided additional details.

A simple theoretical explanation of these observations has been given by Argaman.[44] In weak links, the Josephson critical current depends on the energy distribution of the electrons in the coupling medium. Argaman points out that, under the bias conditions needed for the observation of Shapiro steps, the energy distribution function of the electrons in the semiconductor contains a component that oscillates in time with the Josephson frequency, even in the absence of an external high-frequency drive signal. We must then expect that, in the fundamental relation for the a.c. Josephson current under voltage bias,

$$I(t) = I_J \cdot \sin \omega_J t \quad (\omega_J = 2eV/\hbar), \quad (8)$$

the critical current $I_J$ contains itself a component oscillating in time with the Josephson frequency. If, for simplicity, we assume that the oscillatory component of $I_J$ is cosine-like, we obtain a current contribution at twice the Josephson frequency, of the form

$$I_2(t) \propto \sin 2\omega_J t, \quad (9)$$

which explains the observations of Shapiro steps at half the canonical voltage. Argaman's theory makes several explicit quantitative predictions, almost all of which have been confirmed by the work of Lehnert et al.[43] In particular, the dependence of the half-voltage Shapiro steps on the amplitude of the external high-frequency drive signal shows that the half-voltage steps persist to weak a.c. drive amplitudes and are *not* the result of a non-linearity under conditions of strong a.c. drive.

The idea that the corner currents might contain a non-equilibrium contribution evidently calls for a direct experimental determination of the relative magnitudes of the two contributions. We have performed limited preliminary experiments towards this goal, based on the idea that a cross-over from equilibrium to non-equilibrium might be associated with an abrupt increase in the experimental $R(I)$ characteristic from a near-zero value below the noise floor to a distinctly measurable value above. Measurements on several of our better samples (including sample A), at temperatures below about 4 K, show such transitions, during which $R(I)$ increases by two decades and more over a current range at most a few tens of microampere wide, often much narrower. At sufficiently low temperature, the transition current is typically about one-half the corner current. It exhibits a rapid decrease with increasing temperature, dropping to negligible values above 4.2 K, even though the corner current remains very distinct at those temperatures. The steep rise in $R(I)$ at the transition current is followed by a more gradual rise towards the asymptotic value at the corner current, in a way that varies considerably from sample to sample.

Although these observations are *qualitatively* in line with what the non-equilibrium arguments suggests, they cannot, by themselves, be considered proof of that argument. In fact, if one ignores this argument, alternative explanations suggest themselves. In particular, the data could be fitted by the ad-hoc assumption of significant cell-to-cell variations within each array. Such variations would also explain the observed pronounced sample dependence of the final rise of $R(I)$, including additional complications found in at least some samples, such as asymmetries with current direction, and hysteresis effects. On balance, we therefore believe that these preliminary results, while suggestive of non-equilibrium effects, call for the investigation on single-gap (non-array) samples, with equipment having a lower noise floor. We are, at this time, not able to offer such data.

## V. ACKNOWLEDGMENTS

We have benefited greatly from discussions with many individuals, especially Prof. Reiner Kümmel (Würzburg) and Dr. Nathan Argaman (Santa Barbara). We also wish to express our thanks to Prof. Vinay Ambegaokar (Cornell) for an extensive discussion concerning our observations and the unsuitability of the Ambegaokar-Halperin theory for their explanation.

Our work was initially supported by the Office of Naval research; after cancellation of the ONR support, we were able to complete the work under support from the NSF, through the NSF Science and Technology Center for Quantized Electronic Structures (QUEST), Grant DMR 91-20007.